%% file: main.tex
\documentclass[10pt,twocolumn,letterpaper]{article}

 \usepackage[pagenumbers]{cvpr} 
\usepackage{multirow}
\usepackage{pifont}
\newcommand{\cmark}{\ding{51}}
\newcommand{\xmark}{\ding{55}}

\input{preamble}
\definecolor{cvprblue}{rgb}{0.21,0.49,0.74}
\usepackage[pagebackref,breaklinks,colorlinks,allcolors=cvprblue]{hyperref}

\title{TRITRANS: Spatial-Depth Transformer for 3D Computational Pathology}

\author{Jiali Hu\textsuperscript{1*}, Jiawen Li\textsuperscript{2*}, Yuxuan Chen\textsuperscript{3}, Yonghong He\textsuperscript{2\dag}, Ximing Li\textsuperscript{1\dag}\\
\textsuperscript{1}City University of Hong Kong (Dongguan)\\
\textsuperscript{2}Tsinghua Shenzhen International Graduate School (SIGS), Tsinghua University\\
\textsuperscript{3}The University of Hong Kong\\
{\tt\small heyh@sz.tsinghua.edu.cn}, {\tt\small ximing.li@cityu-dg.edu.cn} \\
}
\begin{document}
\maketitle
\input{sec/0_abstract}    
{\let\thefootnote\relax\footnote{* Contributed equally. \dag Corresponding authors.}}
\input{sec/1_intro}
\input{sec/2_related_work}

\input{sec/3_methodology}

\input{sec/4_experiments}

\input{sec/5_conclusion}
{
    \small
    \bibliographystyle{ieeenat_fullname}
    \bibliography{main}
}

\end{document}

%% file: sec/0_abstract.tex
\begin{abstract}
Three-dimensional (3D) histopathology specimens provide complementary morphological cues across tissue depths. Their effective integration requires capturing both intra-plane spatial relationships and inter-depth dependencies. Existing methods typically extend two-dimensional (2D) histopathology whole-slide image (WSI) models to volumetric data, but lack global interactions across the entire 3D volume. To address this limitation, we introduce TRITRANS (\textbf{Tri}age \textbf{Trans}former), a spatial-depth Transformer network for specimen-level diagnosis in 3D computational pathology. TRITRANS factorizes volumetric contextual modeling into two stages. First, the intra-plane Transformer incorporates spatial positional encodings to model relationships among tissue regions within each optical section. Then the inter-depth Transformer incorporates depth positional encodings to capture dependencies across the entire depth range. Experiments using three pre-trained pathology feature extractors on two public 3D pathology datasets against five baselines demonstrate that TRITRANS achieves state-of-the-art overall performance. Further ablation studies validate the effectiveness of the plane-first modeling order, spatial and depth positional encodings, and Transformer-based interaction strategies. These results highlight the value of modeling spatial and cross-depth context for 3D pathology diagnosis. 

\end{abstract}

%% file: sec/1_intro.tex
\section{Introduction}
\begin{figure}[t]
    \centering
    \includegraphics[width=1\linewidth]{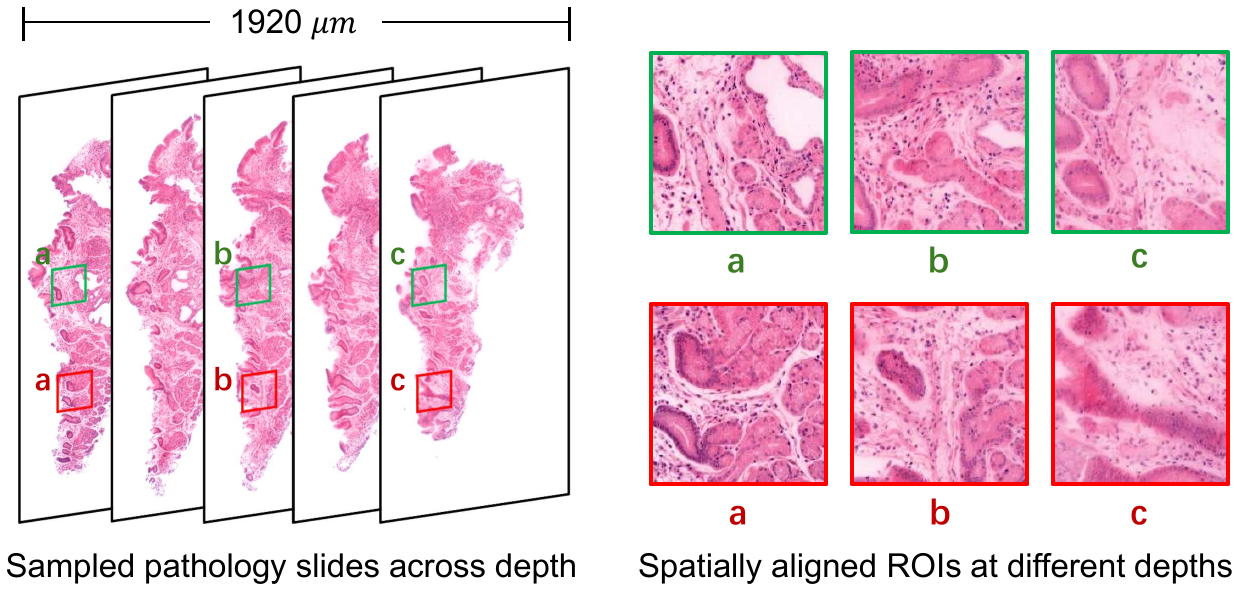}
    \caption{Morphological variations across specimen depth in the 3D histopathology specimen. ROIs refer to regions of interest.}
    \label{fig:fig_1}
\end{figure}

Three-dimensional (3D) histopathology specimens enable volumetric imaging across multiple tissue depths, producing a sequential stack of two-dimensional (2D) optical sections~\cite{liu2021harnessing,yoshikawa2024blueprints,bishop2024end,liu2026ultrarapid}. As shown in Figure~\ref{fig:fig_1}, unlike a single 2D whole-slide image (WSI) that captures only a thin planar cut, optical sections acquired at different depths provide complementary morphological cues and reveal continuous variations in tissue architecture that are often ambiguous in isolated views~\cite{liu2024engineering,song2023artificial,park2025revealing}. Therefore, accurately integrating these volumetric cues for specimen-level diagnosis is a central goal in 3D computational pathology~\cite{song2024analysis}, which requires the model to jointly capture both intra-plane spatial relationships within each optical section and inter-depth dependencies across tissue depths.

Currently, methods for specimen-level diagnosis in 3D computational pathology are primarily extended from established 2D WSI analysis pipelines. In standard 2D WSI modeling frameworks, a tissue slide is partitioned into several patches, whose features are extracted by a pre-trained encoder and aggregated into a slide representation via multiple instance learning (MIL)~\cite{li2021dual,lu2021data}. To adapt this paradigm to 3D specimens, existing approaches typically follow three extension strategies: (1) pooling 2D patches from all optical sections together as a single bag~\cite{ilse2018attention}, (2) encoding each optical section independently and then fusing neighboring sections~\cite{gao2024triage,gao2026deep,yan2026prototype}, or (3) dividing the volume into local 3D patches and aggregating their features via MIL~\cite{song2024analysis,redekop2024digital,redekop2025generating}.

Despite this progress, jointly modeling intra-plane spatial context and global cross-depth dependencies remains challenging. Unstructured pooling omits explicit depth information, neighboring-section fusion limits cross-depth context to local windows, and local 3D patch encoders capture volumetric structure within individual sub-volumes rather than explicitly modeling relationships between them. The latter approaches also commonly rely on generic 3D backbones \cite{liu2022video,tran2018closer}, without benefiting from pathology-specific pretraining. These limitations motivate structured interaction modeling across both spatial and depth dimensions.

Transformer architectures~\cite{vaswani2017attention} offer a natural solution for modeling global interactions in 3D specimens. However, directly applying self-attention across all patches in a 3D volume is computationally prohibitive and lacks explicit structural and positional priors. To overcome these challenges, we introduce TRITRANS, a spatial-depth Transformer network for specimen-level diagnosis in 3D computational pathology. Specifically, TRITRANS first treats patch features as tokens with 2D spatial positional encodings and employs intra-plane Transformer blocks to model relationships among tissue regions, thereby generating a plane token for each optical section. Subsequently, TRITRANS incorporates depth positional encodings into these plane tokens and applies inter-depth Transformer blocks to capture dependencies across the entire depth range, obtaining a task-adaptive specimen token for classification. To evaluate TRITRANS, we conduct experiments on two public 3D pathology datasets for esophageal cancer screening and prostate risk assessment~\cite{gao2026deep}. Using three widely adopted pre-trained 2D pathology feature extractors~\cite{lu2024visual,filiot2025distilling,chen2024towards}, we compare TRITRANS against five representative baselines. Our results demonstrate that TRITRANS achieves state-of-the-art overall performance, outperforming the second-best method in mean balanced accuracy by 5.50\% and 5.37\% on the esophageal and prostate datasets, respectively. Meanwhile, we conduct local 2.5D evaluations and extensive ablation studies on the modeling order, positional encodings, interaction strategies, and network depth to further demonstrate the effectiveness of TRITRANS. We believe TRITRANS will advance accurate specimen-level diagnosis in 3D pathology.

%% file: sec/2_related_work.tex
\section{Related works}
\subsection{3D Computational Pathology}
Recent advances in volumetric histopathology imaging, such as open-top light-sheet microscopy~\cite{glaser2017light}, have enabled multi-depth tissue examination by capturing the 3D morphological structure of specimens across sequential optical sections. Since manually inspecting hundreds of high-resolution optical sections per specimen is prohibitively labor-intensive, integrating artificial intelligence (AI) with volumetric microscopy has driven the emergence of 3D computational pathology for automated structural quantification and clinical diagnosis~\cite{liu2021harnessing,liu2024engineering}. Early studies primarily combined deep learning methods with handcrafted 3D morphological features, such as quantifying 3D glandular architecture~\cite{xie2022prostate} or analyzing volumetric nuclear features~\cite{serafin2023nondestructive} for prostate cancer risk stratification and assessment. To bypass explicit tissue segmentation and handcrafted feature engineering, recent frameworks have shifted toward end-to-end representation learning directly from volumetric data. For instance, TRIPATH~\cite{song2024analysis} employs 3D encoders to extract local volumetric patch features for attention-based MIL prediction, while TRICARE~\cite{gao2026deep} uses attention pooling to integrate neighboring optical sections in a 2.5D-style manner for section-level risk triage in 3D pathology volumes. In addition, recent studies have explored graph-based~\cite{zhao2026scalable} and prototype-based~\cite{yan2026prototype} methods to further enhance 3D structural representation and diagnostic assessment. However, these methods typically focus on local modeling, limiting the ability to capture global interactions across the entire 3D volume.

\subsection{Pathology Foundation Models}
To obtain powerful pathology-specific representations, large-scale 2D pathology foundation models have rapidly emerged through diverse pre-training paradigms. By learning from large-scale histopathology images across diverse tissue types, these models serve as general-purpose pathology encoders capable of extracting rich, high-dimensional feature representations from 2D patches. For example, self-supervised visual encoders such as UNI~\cite{chen2024towards}, Virchow~\cite{vorontsov2024foundation}, and Prov-GigaPath~\cite{xu2024whole} learn generalizable histological features from massive unannotated patch datasets, while vision--language encoders such as PLIP~\cite{huang2023visual}, CONCH~\cite{lu2024visual}, KEEP~\cite{zhou2026knowledge}, and MUSK~\cite{xiang2025vision} incorporate image--text supervision to support both visual recognition and cross-modal applications. In addition, knowledge distillation-based encoders such as H0-mini~\cite{filiot2025distilling}, GPFM~\cite{ma2026generalizable}, and ALICE~\cite{li2026alice} have recently been developed to further enhance representation or enable lightweight deployment. Despite these breakthroughs in 2D histology, foundation models for 3D pathology remain scarce due to the limited scale of volumetric cohorts. Consequently, effectively adapting pre-trained 2D patch representations to 3D specimen-level diagnosis critically depends on downstream modeling frameworks.

\begin{figure*}[t]
    \centering
    \includegraphics[width=1\linewidth]{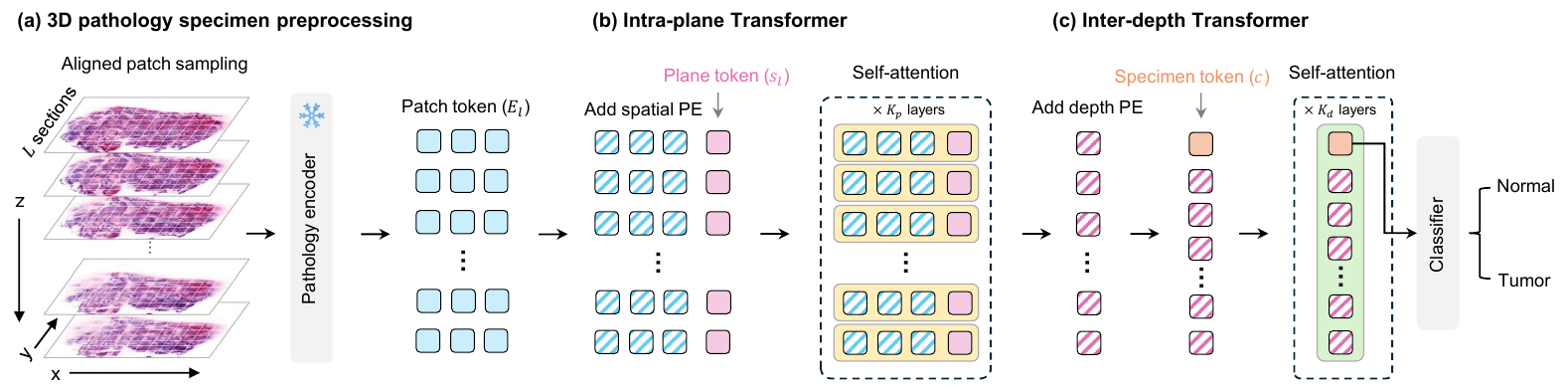}
    \caption{Overview of our TRITRANS model. First, spatially aligned patches from multi-depth sections are encoded by a frozen pathology encoder. Then, the intra-plane Transformer aggregates patch features with spatial positional encoding (PE) into the section-level plane token. Finally, the inter-depth Transformer fuses plane tokens across depth with depth positional encoding into a specimen token for specimen-level prediction.}
    \label{fig:fig_2}
\end{figure*}

\subsection{Multiple Instance Learning for Pathology}
With advances in pathology foundation models, MIL has rapidly evolved into a standard paradigm in 2D computational pathology. By formulating each WSI as a bag of instances, MIL can aggregate patch-level features into a slide representation under weak supervision~\cite{neidlinger2025benchmarking,xu2025multiple,luo2026nnmil,gao2025smmile}. For example, attention-based methods learn adaptive instance weights or multi-branch attention mechanisms to highlight diagnostically informative regions for bag-level pooling~\cite{ilse2018attention,lu2021data,li2021dual,zhang2024attention}. Graph-based methods construct spatial or embedding graphs over patches to capture topological relationships within the tissue microenvironment~\cite{zheng2022graph,li2024dynamic,chen2021whole}. Transformer-based methods use self-attention mechanisms to model correlations across patch tokens throughout a plane-level slide~\cite{shao2021transmil,li2024rethinking}. In 3D computational pathology, these MIL frameworks can be adapted by collecting tiles from all optical sections into a bag set. However, such adaptations ignore volumetric structure and depth location modeling, making it difficult to capture global interactions across the entire 3D volume.

\label{sec:related_works}

%% file: sec/3_methodology.tex
\section{Methodology}
To achieve diagnosis from 3D histopathology specimens, TRITRANS sequentially employs the intra-plane Transformer to model spatial relationships within each optical section and the inter-depth Transformer to capture dependencies across tissue depths. The overall framework of TRITRANS is illustrated in Figure~\ref{fig:fig_2}.

\subsection{Intra-Plane Transformer}
Given a 3D histopathology specimen consisting of $L$ optical sections sampled at imaging depth levels $\mathbf{z} = [z_1, \dots, z_L]^\top \in \mathbb{R}^L$, we first use the Otsu thresholding method to construct a minimal union tissue mask across all sections. For each section $l \in \{1, \dots, L\}$, $N$ non-overlapping patches with a stride of $u$ pixels are then cropped within this union mask at 2D coordinates $\{(x_{l,n}, y_{l,n})\}_{n=1}^N$ and encoded by a frozen pre-trained 2D pathology encoder into patch features $\mathbf{X}_l \in \mathbb{R}^{N \times C}$, where $C$ is the feature dimension. Next, we use a projection layer to map $\mathbf{X}_l$ into patch tokens $\mathbf{E}_l = [\mathbf{e}_{l,1},\, \dots,\, \mathbf{e}_{l,N}] \in \mathbb{R}^{N \times d}$, and add a 2D sinusoidal positional encoding~\cite{vaswani2017attention} $\phi_d(p) \in \mathbb{R}^d$ to each patch token $\mathbf{e}_{l,n} \in \mathbb{R}^d$:
\begin{equation}
    \tilde{\mathbf{e}}_{l,n} = \mathbf{e}_{l,n} + \left[\phi_{d/2}(\bar{x}_{l,n}) \;\|\; \phi_{d/2}(\bar{y}_{l,n})\right],
\end{equation}
\begin{equation}
    \bar{x}_{l,n} = \frac{x_{l,n} - \min_{n'} x_{l,n'}}{u},\quad \bar{y}_{l,n} = \frac{y_{l,n} - \min_{n'} y_{l,n'}}{u}.
\end{equation}
Where $\|$ denotes vector concatenation, $\bar{x}_{l,n}$ and $\bar{y}_{l,n}$ are the normalized grid coordinates.

Subsequently, we introduce the learnable plane token $\mathbf{s}_l$, and form the input sequence $\mathbf{T}_l^{(0)} = [\mathbf{s}_l,\, \tilde{\mathbf{e}}_{l,1},\, \dots,\, \tilde{\mathbf{e}}_{l,N}] \in \mathbb{R}^{(N+1) \times d}$ for section $l$. We then feed $\mathbf{T}_l^{(0)}$ into $K_p$ stacked Transformer blocks to model intra-plane spatial relationships among patches:
\begin{equation}
\begin{aligned}
    \mathbf{R}_l^{(k)} &= \mathbf{T}_l^{(k-1)} + \mathrm{MSA}\!\left(\mathrm{LN}\!\left(\mathbf{T}_l^{(k-1)}\right)\right), \\
    \mathbf{T}_l^{(k)} &= \mathbf{R}_l^{(k)} + \mathrm{FFN}\!\left(\mathrm{LN}\!\left(\mathbf{R}_l^{(k)}\right)\right),
\end{aligned}
\label{eq:intra_block}
\end{equation}
where $\mathrm{MSA}$, $\mathrm{FFN}$, and $\mathrm{LN}$ denote multi-head self-attention, a feed-forward network, and layer normalization, respectively. After the $K_p$-th block, we extract the updated plane token $\mathbf{h}_l = \mathbf{T}_{l,0}^{(K_p)} \in \mathbb{R}^d$ as the representation of section $l$, and stack the plane tokens from all $L$ sections into $\mathbf{H} = [\mathbf{h}_1;\, \dots;\, \mathbf{h}_L] \in \mathbb{R}^{L \times d}$.

\subsection{Inter-Depth Transformer}
After obtaining the plane tokens, the inter-depth Transformer further models dependencies across all $L$ optical sections. To incorporate the depth position of each section, we add a 1D sinusoidal positional encoding to each plane token $\mathbf{h}_l$ based on its imaging depth level $z_l$:
\begin{equation}
    \tilde{\mathbf{h}}_l = \mathbf{h}_l + \phi_d(\bar{z}_l),\quad \bar{z}_l = z_l - \min_{l'} z_{l'},
\end{equation}
where $\bar{z}_l$ denotes the relative depth level with respect to the shallowest section.

Then, we introduce a learnable specimen token $\mathbf{c} \in \mathbb{R}^d$, and form the input sequence $\mathbf{V}^{(0)} = [\mathbf{c},\, \tilde{\mathbf{h}}_1,\, \dots,\, \tilde{\mathbf{h}}_L] \in \mathbb{R}^{(L+1) \times d}$. We then feed $\mathbf{V}^{(0)}$ into $K_d$ stacked Transformer blocks to capture inter-depth dependencies across all sections:
\begin{equation}
\begin{aligned}
    \mathbf{U}^{(k)} &= \mathbf{V}^{(k-1)} + \mathrm{MSA}\!\left(\mathrm{LN}\!\left(\mathbf{V}^{(k-1)}\right)\right), \\
    \mathbf{V}^{(k)} &= \mathbf{U}^{(k)} + \mathrm{FFN}\!\left(\mathrm{LN}\!\left(\mathbf{U}^{(k)}\right)\right).
\end{aligned}
\label{eq:inter_block}
\end{equation}
After the $K_d$-th block, we extract the updated specimen token $\mathbf{g} = \mathbf{V}_0^{(K_d)} \in \mathbb{R}^d$ as the final specimen representation, and feed it into a linear classifier to obtain the specimen-level prediction logits.

\begin{table*}[tbp]
\centering
\resizebox{2\columnwidth}{!}{
\renewcommand{\arraystretch}{1.5}{
\begin{tabular}{llcccccc}
\hline
\multirow{2}{*}{\textbf{Encoders}} &
  \multirow{2}{*}{\textbf{Methods}} &
  \multicolumn{3}{c}{\textbf{Esophagus}} &
  \multicolumn{3}{c}{\textbf{Prostate}} \\ \cline{3-8} 
 &
   &
  \textbf{Bal acc.} &
  \textbf{\textit{w}F2} &
  \multicolumn{1}{c|}{\textbf{AUC}} &
  \textbf{Bal acc.} &
  \textbf{\textit{w}F2} &
  \textbf{AUC} \\ \hline
  
Swin3D-B~\cite{liu2022video} &
  \multirow{2}{*}{TRIPATH~\cite{song2024analysis}} &
  $0.570_{0.032}$ &
  $0.581_{0.036}$ &
  \multicolumn{1}{c|}{$0.564_{0.054}$} &
  $0.647_{0.069}$ &
  $0.522_{0.163}$ &
  $0.761_{0.014}$ \\
R(2+1)D~\cite{tran2018closer} &
         &
  $0.473_{0.033}$ &
  $0.473_{0.049}$ &
  \multicolumn{1}{c|}{$0.470_{0.031}$} &
  $0.584_{0.049}$ &
  $0.331_{0.176}$ &
  $0.746_{0.049}$ \\ \hline

\multirow{5}{*}{H0-mini~\cite{filiot2025distilling}} &
  MeanMIL 2D &
  $0.508_{0.015}$ &
  $0.501_{0.018}$ &
  \multicolumn{1}{c|}{$0.652_{0.018}$} &
  $0.522_{0.046}$ &
  $0.320_{0.181}$ &
  $0.716_{0.020}$ \\

 &
  ABMIL 2D~\cite{ilse2018attention} &
  $0.694_{0.071}$ &
  $0.709_{0.073}$ &
  \multicolumn{1}{c|}{$0.770_{0.106}$} &
  $0.657_{0.053}$ &
  $0.798_{0.042}$ &
  $0.722_{0.019}$ \\

 &
  TransMIL 2D~\cite{shao2021transmil} &
  $0.556_{0.057}$ &
  $0.558_{0.065}$ &
  \multicolumn{1}{c|}{$0.641_{0.151}$} &
  $0.688_{0.057}$ &
  $0.779_{0.065}$ &
  $0.754_{0.029}$ \\

 &
  TRICARE~\cite{gao2026deep} &
  \multicolumn{1}{c}{$0.592_{0.087}$} &
  \multicolumn{1}{c}{$0.596_{0.098}$} &
  \multicolumn{1}{c|}{$0.651_{0.112}$} &
  \multicolumn{1}{c}{$0.645_{0.094}$} &
  \multicolumn{1}{c}{$0.813_{0.044}$} &
  \multicolumn{1}{c}{$0.722_{0.058}$} \\ \cline{2-8}

 &
  \textbf{TRITRANS (Ours)} &
  \multicolumn{1}{c}{$\mathbf{0.754_{0.057}}$} &
  \multicolumn{1}{c}{$\mathbf{0.760_{0.047}}$} &
  \multicolumn{1}{c|}{$\mathbf{0.863_{0.035}}$} &
  \multicolumn{1}{c}{$\mathbf{0.707_{0.067}}$} &
  \multicolumn{1}{c}{$\mathbf{0.820_{0.036}}$} &
  \multicolumn{1}{c}{$\mathbf{0.764_{0.043}}$} \\ \hline

\multirow{5}{*}{UNI~\cite{chen2024towards}} &
  MeanMIL 2D &
  $0.546_{0.057}$ &
  $0.545_{0.064}$ &
  \multicolumn{1}{c|}{$0.691_{0.059}$} &
  $0.590_{0.068}$ &
  $0.538_{0.159}$ &
  $0.603_{0.101}$ \\

 &
  ABMIL 2D~\cite{ilse2018attention} &
  $0.657_{0.049}$ &
  $0.671_{0.053}$ &
  \multicolumn{1}{c|}{$0.796_{0.078}$} &
  $0.678_{0.033}$ &
  $0.729_{0.021}$ &
  $0.733_{0.021}$ \\

 &
  TransMIL 2D~\cite{shao2021transmil} &
  $0.634_{0.055}$ &
  $0.647_{0.058}$ &
  \multicolumn{1}{c|}{$0.622_{0.041}$} &
  $0.620_{0.070}$ &
  $0.589_{0.159}$ &
  $0.696_{0.024}$ \\

 &
  TRICARE~\cite{gao2026deep} &
  \multicolumn{1}{c}{$0.669_{0.056}$} &
  \multicolumn{1}{c}{$0.683_{0.060}$} &
  \multicolumn{1}{c|}{$0.725_{0.074}$} &
  \multicolumn{1}{c}{$0.673_{0.061}$} &
  \multicolumn{1}{c}{$0.700_{0.101}$} &
  \multicolumn{1}{c}{$0.674_{0.050}$} \\ \cline{2-8}

 &
  \textbf{TRITRANS (Ours)} &
  \multicolumn{1}{c}{$\mathbf{0.717_{0.049}}$} &
  \multicolumn{1}{c}{$\mathbf{0.731_{0.048}}$} &
  \multicolumn{1}{c|}{$\mathbf{0.821_{0.039}}$} &
  \multicolumn{1}{c}{$\mathbf{0.688_{0.020}}$} &
  \multicolumn{1}{c}{$\mathbf{0.761_{0.036}}$} &
  \multicolumn{1}{c}{$\mathbf{0.777_{0.038}}$} \\ \hline

\multirow{5}{*}{CONCH~\cite{lu2024visual}} &
  MeanMIL 2D &
  $0.500_{0.000}$ &
  $0.491_{0.000}$ &
  \multicolumn{1}{c|}{$0.709_{0.030}$} &
  $0.509_{0.017}$ &
  $0.135_{0.036}$ &
  $0.581_{0.038}$ \\

 &
  ABMIL 2D~\cite{ilse2018attention} &
  $0.678_{0.064}$ &
  $0.693_{0.065}$ &
  \multicolumn{1}{c|}{$0.757_{0.102}$} &
  $0.559_{0.027}$ &
  $0.470_{0.087}$ &
  $0.696_{0.010}$ \\

 &
  TransMIL 2D~\cite{shao2021transmil} &
  $0.588_{0.033}$ &
  $0.595_{0.039}$ &
  \multicolumn{1}{c|}{$0.672_{0.133}$} &
  $0.616_{0.085}$ &
  $0.587_{0.142}$ &
  $0.642_{0.012}$ \\

 &
  TRICARE~\cite{gao2026deep} &
  \multicolumn{1}{c}{$0.605_{0.045}$} &
  \multicolumn{1}{c}{$0.615_{0.050}$} &
  \multicolumn{1}{c|}{$0.626_{0.119}$} &
  \multicolumn{1}{c}{$0.651_{0.088}$} &
  \multicolumn{1}{c}{$0.750_{0.116}$} &
  \multicolumn{1}{c}{$0.687_{0.010}$} \\ \cline{2-8}

 & 
  \textbf{TRITRANS (Ours)} &
  \multicolumn{1}{c}{$\mathbf{0.723_{0.091}}$} &
  \multicolumn{1}{c}{$\mathbf{0.734_{0.092}}$} &
  \multicolumn{1}{c|}{$\mathbf{0.874_{0.039}}$} &
  \multicolumn{1}{c}{$\mathbf{0.735_{0.011}}$} &
  \multicolumn{1}{c}{$\mathbf{0.813_{0.051}}$} &
  \multicolumn{1}{c}{$\mathbf{0.719_{0.020}}$} \\ \hline
\end{tabular}}}
\caption{Comparison with baseline methods for 3D specimen-level classification on the esophageal and prostate datasets. Results are reported as $\text{mean}_{\text{std}}$ across five folds, with the best performance under each encoder in \textbf{bold}.}
\label{main_table_3d}
\end{table*}

%% file: sec/4_experiments.tex
\section{Experiments}
\subsection{Experimental Setup}


\begin{table*}[t]
\centering
\resizebox{2\columnwidth}{!}{
\renewcommand{\arraystretch}{1.5}{
\begin{tabular}{llcccccc}
\hline
\multirow{2}{*}{\textbf{Encoders}} &
  \multirow{2}{*}{\textbf{Methods}} &
  \multicolumn{3}{c}{\textbf{Esophagus}} &
  \multicolumn{3}{c}{\textbf{Prostate}} \\ \cline{3-8} 
 &
   &
  \textbf{Bal acc.} &
  \textbf{\textit{w}F2} &
  \multicolumn{1}{c|}{\textbf{AUC}} &
  \textbf{Bal acc.} &
  \textbf{\textit{w}F2} &
  \textbf{AUC} \\ \hline

\multirow{5}{*}{H0-mini\cite{filiot2025distilling}} &
  MeanMIL 2D &
  $0.588_{0.050}$ &
  $0.491_{0.080}$ &
  \multicolumn{1}{c|}{$0.675_{0.047}$} &
  $0.524_{0.023}$ &
  $0.259_{0.106}$ &
  $0.718_{0.018}$ \\

 &
  ABMIL 2D~\cite{ilse2018attention} &
  $0.652_{0.043}$ &
  $0.583_{0.068}$ &
  \multicolumn{1}{c|}{$0.781_{0.074}$} &
  $0.679_{0.063}$ &
  $0.811_{0.029}$ &
  $0.733_{0.017}$ \\

 &
  TransMIL 2D~\cite{shao2021transmil} &
  $0.599_{0.064}$ &
  $0.522_{0.073}$ &
  \multicolumn{1}{c|}{$0.667_{0.086}$} &
  $0.661_{0.014}$ &
  $0.750_{0.056}$ &
  $0.707_{0.016}$ \\

 &
  TRICARE~\cite{gao2026deep} &
  \multicolumn{1}{c}{$0.632_{0.051}$} &
  \multicolumn{1}{c}{$0.559_{0.076}$} &
  \multicolumn{1}{c|}{$0.720_{0.068}$} &
  \multicolumn{1}{c}{$0.608_{0.040}$} &
  \multicolumn{1}{c}{$0.808_{0.024}$} &
  \multicolumn{1}{c}{$0.701_{0.011}$} \\ \cline{2-8}

 &
  \textbf{TRITRANS (Ours)} &
  \multicolumn{1}{c}{$\mathbf{0.670_{0.042}}$} &
  \multicolumn{1}{c}{$\mathbf{0.616_{0.052}}$} &
  \multicolumn{1}{c|}{$\mathbf{0.786_{0.041}}$} &
  \multicolumn{1}{c}{$\mathbf{0.691_{0.067}}$} &
  \multicolumn{1}{c}{$\mathbf{0.832_{0.022}}$} &
  \multicolumn{1}{c}{$\mathbf{0.758_{0.025}}$} \\ \hline

\multirow{5}{*}{UNI~\cite{chen2024towards}} &
  MeanMIL 2D &
  $0.590_{0.051}$ &
  $0.495_{0.079}$ &
  \multicolumn{1}{c|}{$0.691_{0.042}$} &
  $0.580_{0.062}$ &
  $0.432_{0.134}$ &
  $0.566_{0.088}$ \\

 &
  ABMIL 2D~\cite{ilse2018attention} &
  $0.653_{0.010}$ &
  $0.582_{0.012}$ &
  \multicolumn{1}{c|}{$0.798_{0.035}$} &
  $0.630_{0.010}$ &
  $0.600_{0.235}$ &
  $0.670_{0.051}$ \\

 &
  TransMIL 2D~\cite{shao2021transmil} &
  $0.488_{0.017}$ &
  $0.345_{0.010}$ &
  \multicolumn{1}{c|}{$0.482_{0.065}$} &
  $0.630_{0.064}$ &
  $0.599_{0.157}$ &
  $0.691_{0.033}$ \\

 &
  TRICARE~\cite{gao2026deep}&
  \multicolumn{1}{c}{$0.666_{0.036}$} &
  \multicolumn{1}{c}{$0.607_{0.055}$} &
  \multicolumn{1}{c|}{$0.786_{0.014}$} &
  \multicolumn{1}{c}{$0.687_{0.058}$} &
  \multicolumn{1}{c}{$0.710_{0.095}$} &
  \multicolumn{1}{c}{$0.678_{0.011}$} \\ \cline{2-8}

 &
  \textbf{TRITRANS (Ours)} &
  \multicolumn{1}{c}{$\mathbf{0.728_{0.053}}$} &
  \multicolumn{1}{c}{$\mathbf{0.679_{0.066}}$} &
  \multicolumn{1}{c|}{$\mathbf{0.813_{0.042}}$} &
  \multicolumn{1}{c}{$\mathbf{0.707_{0.030}}$} &
  \multicolumn{1}{c}{$\mathbf{0.735_{0.097}}$} &
  \multicolumn{1}{c}{$\mathbf{0.767_{0.054}}$} \\ \hline

\multirow{5}{*}{CONCH~\cite{lu2024visual}} &
  MeanMIL 2D &
  $0.533_{0.032}$ &
  $0.401_{0.047}$ &
  \multicolumn{1}{c|}{$0.732_{0.046}$} &
  $0.501_{0.019}$ &
  $0.149_{0.031}$ &
  $0.626_{0.050}$ \\

 &
  ABMIL 2D~\cite{ilse2018attention}&
  $0.632_{0.037}$ &
  $0.559_{0.059}$ &
  \multicolumn{1}{c|}{$0.799_{0.074}$} &
  $0.542_{0.056}$ &
  $0.406_{0.127}$ &
  $0.673_{0.006}$ \\

 &
  TransMIL~\cite{shao2021transmil}&
  $0.642_{0.044}$ &
  $0.577_{0.070}$ &
  \multicolumn{1}{c|}{$0.678_{0.066}$} &
  $0.680_{0.048}$ &
  $0.692_{0.075}$ &
  $0.662_{0.009}$ \\

 &
  TRICARE~\cite{gao2026deep}&
  \multicolumn{1}{c}{$0.595_{0.035}$} &
  \multicolumn{1}{c}{$0.502_{0.054}$} &
  \multicolumn{1}{c|}{$0.772_{0.077}$} &
  \multicolumn{1}{c}{$0.667_{0.046}$} &
  \multicolumn{1}{c}{$0.707_{0.089}$} &
  \multicolumn{1}{c}{$0.661_{0.028}$} \\ \cline{2-8}

 & 
  \textbf{TRITRANS (Ours)} &
  \multicolumn{1}{c}{$\mathbf{0.706_{0.084}}$} &
  \multicolumn{1}{c}{$\mathbf{0.659_{0.120}}$} &
  \multicolumn{1}{c|}{$\mathbf{0.801_{0.030}}$} &
  \multicolumn{1}{c}{$\mathbf{0.744_{0.038}}$} &
  \multicolumn{1}{c}{$\mathbf{0.810_{0.062}}$} &
  \multicolumn{1}{c}{$\mathbf{0.690_{0.009}}$} \\ \hline
\end{tabular}}}
\caption{Comparison with baseline methods for 2.5D section-level classification on the esophageal and prostate datasets. Results are reported as $\text{mean}_{\text{std}}$ across five folds, with the best performance under each encoder in \textbf{bold}.}
\label{main_table_2.5d}
\end{table*}

\subsubsection{Datasets and Tasks}
We conduct experiments on two public H\&E-stained 3D pathology datasets~\cite{gao2026deep} for esophageal cancer screening and prostate risk assessment. The esophageal dataset consists of a development cohort of 103 specimens from 24 patients (21 tumor, 82 normal; 6,978 optical sections) and a held-out test cohort of 30 specimens from 5 patients (13 tumor, 17 normal; 231 optical sections). The prostate dataset comprises a development cohort of 121 specimens from 54 patients (68 tumor, 53 normal; 3,993 optical sections) and a held-out test cohort of 29 specimens from 29 patients (6 tumor, 23 normal; 261 optical sections). Our primary task is 3D specimen-level classification, where all optical sections of a specimen are integrated to predict a single tumor-versus-normal specimen label. In addition, to evaluate local cross-depth modeling, we also conduct a 2.5D section-level classification experiment following~\cite{gao2026deep}, where each input consists of a target optical section and one adjacent section on either side to predict the target section's label. For all experiments, we perform 5-fold patient-level cross-validation on the development cohort and evaluate all five fold-specific models on the held-out test cohort, reporting the mean and standard deviation across the five folds.

\subsubsection{Implementation Details}
We crop non-overlapping $224 \times 224$ patches ($u = 224$) and extract patch features using three frozen pre-trained 2D pathology encoders: H0-mini~\cite{filiot2025distilling}, UNI~\cite{chen2024towards}, and CONCH~\cite{lu2024visual}. For TRITRANS, we set the hidden dimension $d = 512$ and the number of Transformer blocks to $K_p = K_d = 2$, with 8 attention heads and an FFN expansion ratio of 4 in each block. Models are trained for up to 50 epochs using the AdamW optimizer with a weight decay of $10^{-5}$, a batch size of 1 specimen, and early stopping with a patience of 10 epochs based on validation balanced accuracy. The learning rate is set to $10^{-4}$ for the esophageal dataset and $10^{-5}$ for the prostate dataset. Following~\cite{gao2026deep}, to account for class imbalance, we adopt balanced accuracy (Bal acc.), weighted F2 score (\textit{w}F2), and AUC as our evaluation metrics.

\subsubsection{Compared Methods}
We compare TRITRANS against three baselines extended from conventional 2D WSI methods, including MeanMIL 2D, ABMIL 2D~\cite{ilse2018attention}, and TransMIL 2D~\cite{shao2021transmil}, as well as two 3D pathology methods, including TRICARE~\cite{gao2026deep} and TRIPATH~\cite{song2024analysis}. Specifically, MeanMIL 2D globally averages patch features across all optical sections. ABMIL 2D and TransMIL 2D pool patch features from all sections into a single bag for gated-attention or Transformer-based aggregation. TRICARE employs a two-level gated-attention hierarchy to first aggregate patches into section representations and then fuse sections across depths. TRIPATH employs 3D encoders (Swin3D-B~\cite{liu2022video} and R(2+1)D~\cite{tran2018closer}) to extract features from local volumetric patches and aggregates them via gated-attention MIL for 3D specimen-level classification.

\subsection{Comparison with Existing Methods}
\subsubsection{3D Specimen-Level Classification}
Table~\ref{main_table_3d} presents the 3D specimen-level classification results on the esophageal and prostate datasets. TRITRANS outperforms the second-best method in mean balanced accuracy across the three encoders by 5.50\% on the esophageal dataset and 5.37\% on the prostate dataset. Specifically, on the esophageal dataset, TRITRANS reaches a balanced accuracy of 0.754 with H0-mini and an AUC of 0.874 with CONCH, surpassing the strongest baseline by 6.0\% and 11.7\%, respectively. On the prostate dataset with CONCH, TRITRANS achieves a balanced accuracy of 0.735 and a \textit{w}F2 of 0.813, outperforming the second-best TRICARE by 8.4\% and 6.3\%, respectively. Moreover, we observe that TRIPATH with 3D encoders remains competitive on the prostate dataset by directly modeling local 3D structure. This indicates that while pre-trained 2D pathology encoders offer powerful domain-specific representations, they inherently lack 3D spatial context across tissue depths; TRITRANS effectively bridges this gap through spatial-depth contextual modeling, consistently outperforming TRIPATH on both datasets.

\subsubsection{2.5D Section-Level Classification} 
Table~\ref{main_table_2.5d} reports the results for the local 2.5D section-level classification task. Across the three encoders, TRITRANS consistently achieves the best performance on all metrics, outperforming the second-best method in mean balanced accuracy by 5.57\% on the esophageal dataset and 5.70\% on the prostate dataset. Specifically, on the esophageal dataset with UNI, TRITRANS reaches a balanced accuracy of 0.728 and a \textit{w}F2 of 0.679, surpassing the second-best TRICARE by 6.2\% and 7.2\%, respectively. On the prostate dataset with CONCH, TRITRANS achieves a balanced accuracy of 0.744 and a wF2 of 0.810, outperforming TRICARE by 7.7\% and 10.3\%, respectively. Notably, although TRICARE is specifically designed for 2.5D section-level triage via attention pooling, TRITRANS consistently surpasses it across all settings, demonstrating that our spatial-depth Transformer captures local cross-depth cues more effectively even within a narrow depth window.

\subsection{Ablation Studies}
\subsubsection{Effect of Modeling Order}

We first investigate the order of spatial and depth modeling in Table~\ref{tab:ablation_order} by comparing our plane-first order ($\text{L}\to\text{D}$) against the reversed depth-first order ($\text{D}\to\text{L}$). Specifically, in the $\text{D}\to\text{L}$ variant, the projected patch tokens are first added with 1D depth positional encodings and grouped by their 2D grid locations into $N$ sequences of length $L$; each sequence is processed by the inter-depth Transformer with a learnable token to produce a position token, and the resulting $N$ position tokens are then added with 2D spatial positional encodings and fed into the intra-plane Transformer with a specimen token for final classification. Across all three encoders, our $\text{L}\to\text{D}$ order consistently outperforms $\text{D}\to\text{L}$, improving the mean balanced accuracy by 7.63\% on the esophageal dataset and 4.60\% on the prostate dataset. These results indicate that constructing spatially contextualized section-level representations first provides a more reliable foundation for subsequent cross-depth integration.

\subsubsection{Effect of Positional Encodings}

We next evaluate the effectiveness of the spatial positional encoding and the depth positional encoding across all three pathology encoders in Table~\ref{tab:ablation_pe}. Across various combinations of encoders and datasets, the highest balanced accuracy is achieved by simultaneously introducing spatial and depth positional encodings. Specifically, compared with the variant without positional encodings, adding both the spatial and depth PEs improves the mean balanced accuracy across the three encoders by 4.97\% on the esophageal dataset and 4.47\% on the prostate dataset. Furthermore, removing either 2D PE or Depth PE from the full model consistently degrades performance, reducing the mean balanced accuracy by 5.97\% and 6.17\% on the esophageal dataset, and by 4.13\% and 2.83\% on the prostate dataset, respectively. These results confirm that both intra-plane spatial information and inter-depth relationships are essential for 3D contextual modeling.

\subsubsection{Effect of Transformer-Based Interaction}

\begin{figure}[tbp]
    \centering
    \includegraphics[width=1\linewidth]{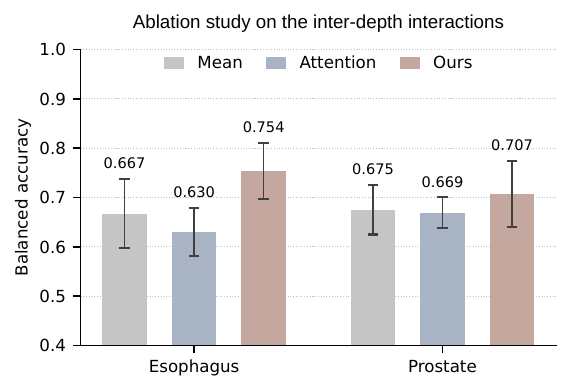}
    \caption{Ablation study on the inter-depth interaction strategy using H0-mini features, where the inter-depth Transformer is replaced with mean or gated-attention pooling. We report balanced accuracy across five folds.}
    \label{fig:ablation_agg_inter}
\end{figure}

\begin{figure}[tbp]
    \centering
    \includegraphics[width=1\linewidth]{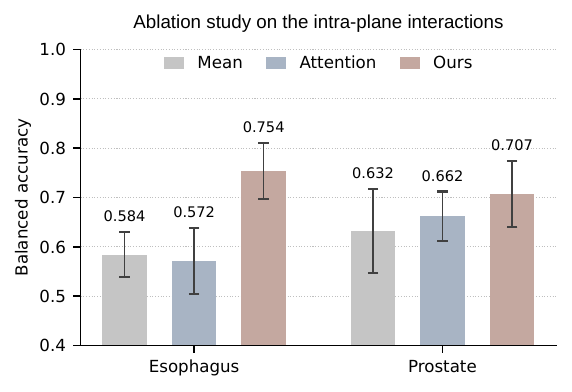}
    \caption{Ablation study on the intra-plane interaction strategy using H0-mini features, where the intra-plane Transformer is replaced with mean or gated-attention pooling. We report balanced accuracy across five folds.}
    \label{fig:ablation_agg_intra}
\end{figure}

To verify the necessity of Transformer-based interactions at both stages, we replace either the inter-depth Transformer or the intra-plane Transformer with mean pooling or attention pooling while keeping the other stage unchanged. As shown in Figure~\ref{fig:ablation_agg_inter}, replacing the inter-depth Transformer with mean or gated-attention pooling reduces the balanced accuracy with H0-mini from 0.754 to 0.667 and 0.630 on the esophageal dataset, and from 0.707 to 0.675 and 0.669 on the prostate dataset, respectively. Similarly, as shown in Figure~\ref{fig:ablation_agg_intra}, replacing the intra-plane Transformer leads to even more pronounced performance drops, decreasing the balanced accuracy with H0-mini by 17.0\% and 18.2\% on the esophageal dataset, and by 7.5\% and 4.5\% on the prostate dataset compared with mean and gated-attention pooling, respectively. These results demonstrate that simple pooling operations may be insufficient and that explicit self-attention interactions are crucial at both the intra-plane and inter-depth stages.

\begin{table}[t]
\centering
\resizebox{0.88\columnwidth}{!}{
\renewcommand{\arraystretch}{1.5}{
\begin{tabular}{ll|c|c}
\hline
\textbf{Encoders} & \textbf{Order} &
\textbf{Esophagus} & \textbf{Prostate} \\ \hline

\multirow{2}{*}{H0-mini~\cite{filiot2025distilling}}
   & D$\rightarrow$L                & $0.644_{0.083}$ & $0.678_{0.057}$ \\
   & L$\rightarrow$D & $\mathbf{0.754_{0.057}}$ & $\mathbf{0.707_{0.067}}$ \\ \cline{1-4}

\multirow{2}{*}{UNI~\cite{chen2024towards}}
   & D$\rightarrow$L                & $0.651_{0.064}$ & $0.599_{0.020}$ \\
   & L$\rightarrow$D & $\mathbf{0.717_{0.049}}$ & $\mathbf{0.688_{0.020}}$ \\ \cline{1-4}
\multirow{2}{*}{CONCH~\cite{lu2024visual}}
   & D$\rightarrow$L                & $0.670_{0.060}$ & $0.715_{0.058}$ \\
   & L$\rightarrow$D & $\mathbf{0.723_{0.091}}$ & $\mathbf{0.735_{0.011}}$ \\ \hline

\end{tabular}}}
\caption{Comparison results on the modeling order ($\text{L}\to\text{D}$: plane-first; $\text{D}\to\text{L}$: depth-first) in balanced accuracy. Results are reported as $\text{mean}_{\text{std}}$ across five folds, with the best performance under each encoder in \textbf{bold}.}
\label{tab:ablation_order}
\end{table}

\begin{table}[t]
\centering
\resizebox{1\columnwidth}{!}{
\renewcommand{\arraystretch}{1.25}{
\begin{tabular}{lcc|c|c}
\hline
\textbf{Encoders} & \textbf{Spatial PE} & \textbf{Depth PE} &
\textbf{Esophagus} & \textbf{Prostate} \\ \hline
\multirow{4}{*}{H0-mini~\cite{filiot2025distilling}}
 & \xmark & \xmark & $0.703_{0.033}$ & $0.704_{0.050}$ \\
 & \cmark & \xmark & $0.636_{0.067}$ & $0.705_{0.015}$ \\
 & \xmark & \cmark & $0.687_{0.074}$ & $0.705_{0.063}$ \\
 & \cmark & \cmark & $\mathbf{0.754_{0.057}}$ & $\mathbf{0.707_{0.067}}$ \\ \hline
\multirow{4}{*}{UNI~\cite{chen2024towards}}
 & \xmark & \xmark & $0.638_{0.082}$ & $0.643_{0.028}$ \\
 & \cmark & \xmark & $0.695_{0.086}$ & $0.672_{0.049}$ \\
 & \xmark & \cmark & $0.684_{0.019}$ & $0.655_{0.061}$ \\
 & \cmark & \cmark & $\mathbf{0.717_{0.049}}$ & $\mathbf{0.688_{0.020}}$ \\ \hline
\multirow{4}{*}{CONCH~\cite{lu2024visual}}
 & \xmark & \xmark & $0.704_{0.094}$ & $0.649_{0.077}$ \\
 & \cmark & \xmark & $0.678_{0.100}$ & $0.668_{0.047}$ \\
 & \xmark & \cmark & $0.644_{0.099}$ & $0.646_{0.074}$ \\
 & \cmark & \cmark & $\mathbf{0.723_{0.091}}$ & $\mathbf{0.735_{0.011}}$ \\ \hline
\end{tabular}}}
\caption{Comparison results on the spatial positional encoding (PE) and depth positional encoding for 3D specimen-level classification. Balanced accuracy is reported as $\text{mean}_{\text{std}}$ across five folds, with the best results in \textbf{bold}.}
\label{tab:ablation_pe}
\end{table}

\subsubsection{Effect of the Number of Transformer Blocks}

\begin{figure}[tbp]
    \centering
    \includegraphics[width=1\linewidth]{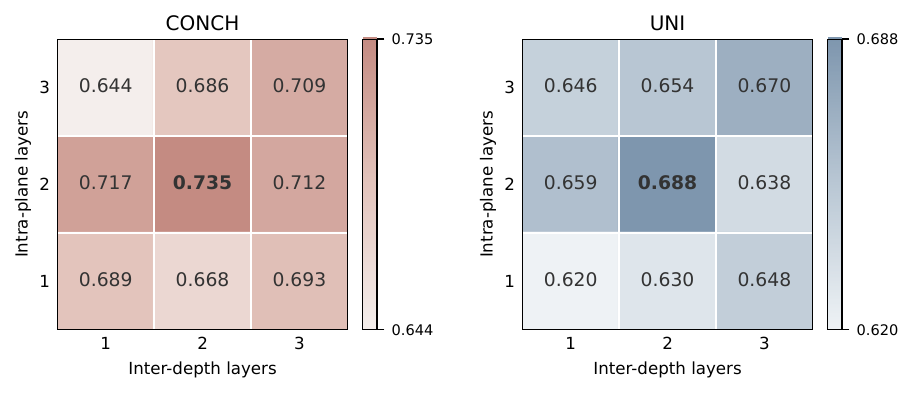}
    \caption{Ablation study on the number of intra-plane ($K_p$) and inter-depth ($K_d$) Transformer blocks on the prostate dataset using CONCH and UNI features. Each cell reports the mean balanced accuracy across five folds.}
    \label{fig:layers_ablation}
\end{figure}

Finally, we investigate the impact of the number of intra-plane Transformer blocks $K_p \in \{1, 2, 3\}$ and inter-depth Transformer blocks $K_d \in \{1, 2, 3\}$ on the prostate dataset using CONCH and UNI features in Figure~\ref{fig:layers_ablation}. For both encoders, our default configuration ($K_p = 2, K_d = 2$) consistently achieves the highest balanced accuracy, reaching 0.735 with CONCH and 0.688 with UNI, and outperforming the second-best configuration by 1.8\% in both cases. Specifically, using a single block ($K_p = 1, K_d = 1$) reduces the balanced accuracy to 0.689 with CONCH and 0.620 with UNI due to insufficient modeling capacity. Conversely, further increasing the depth to $K_p = 3$ and $K_d = 3$ lowers the balanced accuracy to 0.709 and 0.670, respectively, indicating that stacking an excessive number of Transformer blocks may lead to overfitting on limited 3D specimen cohorts.

\subsection{Visualization}

\begin{figure*}[tbp]
    \centering
    \includegraphics[width=1\linewidth]{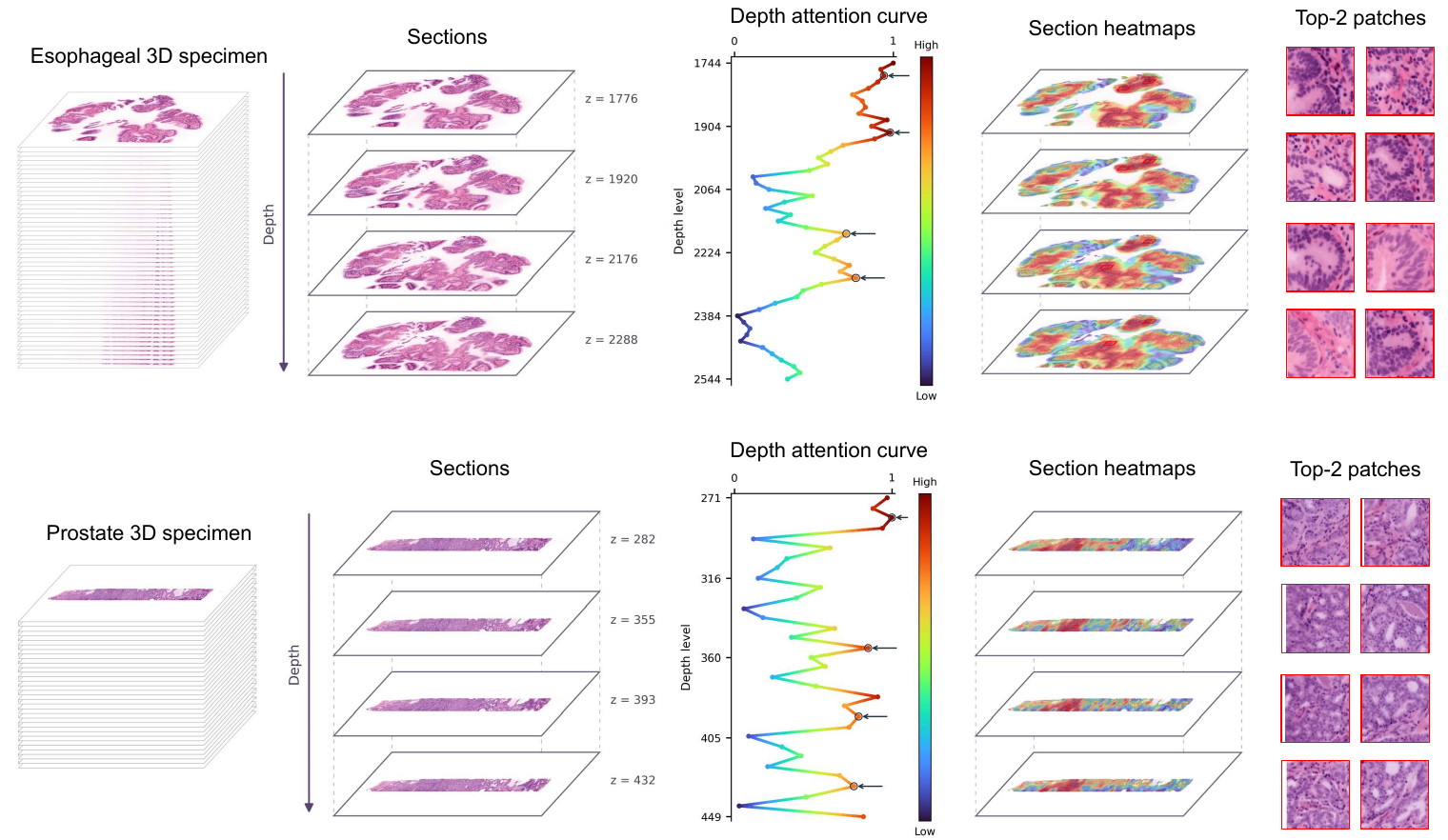}
    \caption{Hierarchical visualization of TRITRANS on representative esophageal (top) and prostate (bottom) specimens. From left to right: the 3D specimen stack, four representative optical sections at selected depth levels $z$, the depth attention curve derived from the inter-depth Transformer (with arrows indicating the selected sections), the spatial attention heatmaps on each section from the intra-plane Transformer, and the corresponding top-2 attended patches.}
    \label{fig:vis_fig}
\end{figure*}

To illustrate the hierarchical interpretability of TRITRANS, Figure~\ref{fig:vis_fig} visualizes the multi-level attention patterns on two representative tumor specimens from the esophageal and prostate datasets. Specifically, inspired by TRICARE~\cite{gao2026deep} and the CLAM codebase~\cite{lu2021data}, from the final block of the inter-depth Transformer, we average the attention weights from the specimen token to all plane tokens across the attention heads and compute their percentile ranks across the specimen, obtaining a relative score in $[0, 1]$ plotted along the imaging depth levels as the depth attention curve. Similarly, from the final block of the intra-plane Transformer, we average the attention weights from the plane token to all patch tokens across heads to select the top-2 patches with the highest attention, and compute within-section percentile ranks followed by Gaussian smoothing to generate the spatial section heatmaps. As shown by the depth attention curves, the relative attention varies markedly across depth levels, indicating that diagnostically informative structures are unevenly distributed along the tissue depth and may be missed in isolated 2D sections. Moreover, the selective section heatmaps can also consistently localize suspicious regions across depths, while the corresponding top-2 patches exhibit typical malignant morphology, such as crowded dysplastic glands and infiltrative neoplastic glands. These visualizations demonstrate that TRITRANS effectively highlights both key optical sections across depth and informative diagnostic regions within each section.

%% file: sec/5_conclusion.tex
\section{Conclusion and Future Work}
In this paper, we introduced TRITRANS, a spatial-depth Transformer network for specimen-level diagnosis in 3D computational pathology. By factorizing volumetric contextual modeling into two sequential stages, TRITRANS first employs an intra-plane Transformer with 2D spatial positional encodings to summarize patch features within each optical section into a plane token, and then applies an inter-depth Transformer with depth positional encodings to capture dependencies across the entire depth range. Extensive experiments on two public 3D pathology datasets across three pre-trained 2D pathology encoders demonstrate that TRITRANS achieves state-of-the-art performance in both 3D specimen-level and local 2.5D section-level classification. Comprehensive ablation studies and multi-level attention visualizations further confirm the effectiveness and hierarchical interpretability of our framework.

In future work, we plan to extend TRITRANS in three directions. First, since each optical section in a 3D specimen inherently exhibits a hierarchical multi-magnification tissue structure, we aim to incorporate multi-scale region modeling into our spatial-depth framework. Second, given the limited scale of 3D pathology cohorts and the strong pre-training capability of Transformer architectures, we will explore transferring knowledge learned from large-scale 2D WSI datasets to pre-train TRITRANS. Finally, to handle ultra-large 3D pathology specimen data with greater depth and larger spatial context, we will investigate computationally efficient sequence modeling mechanisms to further improve the scalability of TRITRANS.